# Top-down fabricated reconfigurable FET with two symmetric and high-current on-states

M. Simon, B. Liang, D. Fischer, M. Knaut, A. Tahn, T. Mikolajick, *Senior Member, IEEE,* and W. M. Weber, *Member, IEEE*

*Abstract*—We demonstrate a top-down fabricated reconfigurable field effect transistor (RFET) based on a silicon nanowire that can be electrostatically programmed to p- and n-configuration. The device unites a high symmetry of transfer characteristics, high on/off current ratios in both configurations and superior current densities in comparison to other top-down fabricated RFETs. Two $NiSi_2$/Si Schottky junctions are formed inside the wire and gated individually. The narrow omega-gated channel is fabricated by a repeated $SiO_2$ etch and growth sequence and a conformal TiN deposition. The gate and Schottky contact metal work functions and the oxide-induced compressive stress to the Schottky junction are adjusted to result in only factor 1.6 higher p- than n-current for in absolute terms identical gate voltages and identical drain voltages.

## I. Introduction

OUR society demands for power-efficient electronic circuits with increasing data throughput and security. As achieving this by simply reducing the device dimensions and operation voltage of field effect transistors (FETs) becomes increasingly difficult, new approaches to generate efficient

This work was supported by DFG in the frameworks of ReproNano (WE 4853/1-3) and the cluster of excellence CfAED (EXC 1056). The authors would like to thank Uwe Mühle (Fraunhofer IKTS, now with Robert Bosch Semiconductor Manufacturing Dresden GmbH, Germany) for TEM and Ricardo Revello (NaMLab, now with Fraunhofer IPMS, 01109 Dresden, Germany) for CV-characterization of TiN.

M. Simon and T. Mikolajick are with NaMLab gGmbH, 01187 Dresden, Germany (e-mail: maik.simon@namlab.com).
B. Liang was with NaMLab. He is now with IMEC, 3001 Leuven, Belgium.
D. Fischer was with the Chair of Semiconductor Technology, Institute of Semiconductors and Microsystems, TU Dresden, 01062 Dresden. He is now with Heliatek GmbH, 01139 Dresden, Germany.
M. Knaut is with the Chair of Semiconductor Technology, Institute of Semiconductors and Microsystems, TU Dresden, 01062 Dresden.
A. Tahn is with the Dresden Center for Nanoanalysis (DCN), 01062 Dresden, Germany.
T. Mikolajick is also with the Chair of Nanoelectronic Materials, TU Dresden, 01187 Dresden, Germany.
M. Simon, A. Tahn and T. Mikolajick are also with the Center for Advancing Electronics Dresden (CfAED), TU Dresden, 01062 Dresden, Germany.
W. M. Weber, was with NaMLab and CfAED. He is now with the Institute of Solid State Electronics, TU Wien, 1040 Vienna, Austria.

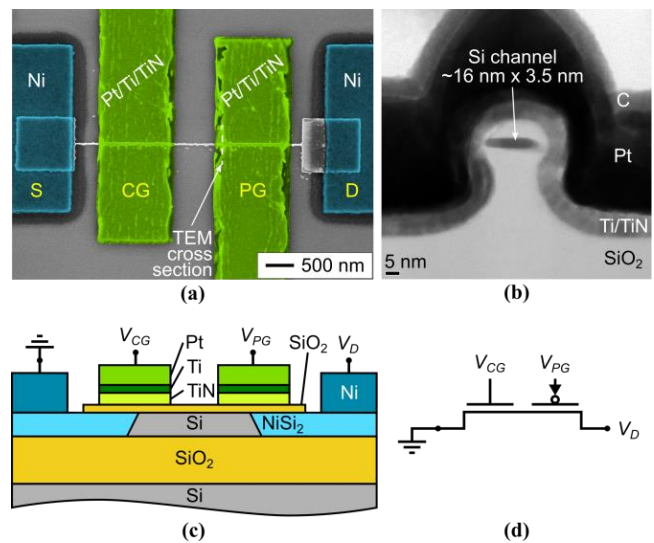

Fig. 1. (a) Colored SEM top-view image of the RFET featuring a control gate (CG) and a program gate (PG). (b) Transmission electron microscopy image of a cut through the PG in (a) showing a nanowire channel with omega-shaped gate stack. Carbon coating originates from cut preparation. (c) Schematic of the transistor in side view along the channel direction. The $NiSi_2$/Si Schottky junctions are covered by the gates. (d) Electronic symbol of the RFET.

logic devices have to be considered. One emerging solution are reconfigurable FETs (RFETs). They need – in contrast to conventional MOSFETs – no physical doping but feature gated Schottky junctions to controllably inject electrons or holes. They can thus be toggled between p- and n-type at runtime. This allows to create compact logic gates that can even change their functionality dynamically, e.g. cells of three RFETs that can work as NAND or NOR [1]–[5]. As a result, the area and structural delay of larger circuits, e.g. of polar decoders, ALUs and adders can be reduced [3]–[5]. Reconfigurability can also strengthen the hardware security of circuits because it obstructs the reverse engineering of their functionality by device imaging or side channel attacks [6].

Several RFETs have been experimentally demonstrated based on bottom-up grown Si or Ge nanowires [7]–[11], carbon nanotubes [12] or two-dimensional materials [13]–[16]. Low band-gap channel materials like Ge are beneficial to increase the current close to CMOS levels [17]. However, a geometrically well-controlled and CMOS-compatible fabrication has so far only been achieved by a top-down fabrication on base of silicon-on-insulator (SOI) wafers or poly-Si films [18]–[30]. Yet, the comparably high band gap of Si demands for high electric fields to induce strong tunneling currents which can be best achieved in narrow channels with a multi-gate architecture. Note that in contrast to Schottky barrier (SB) MOSFETs, the SBs in RFETs cannot be reduced for just one carrier type. Instead, a high symmetry of the IV-characteristics in p- and n-configuration is desired to ensure switching delay indifference. This can be enabled by a precise alignment of gate and Schottky contact metal work functions and stress in the wire. Additionally, RFETs need to be optimized towards high on-currents for fast calculations and low off-currents for a low standby-power consumption.

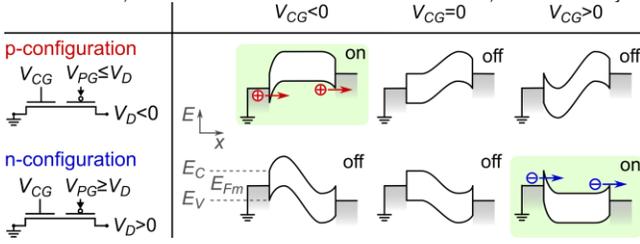

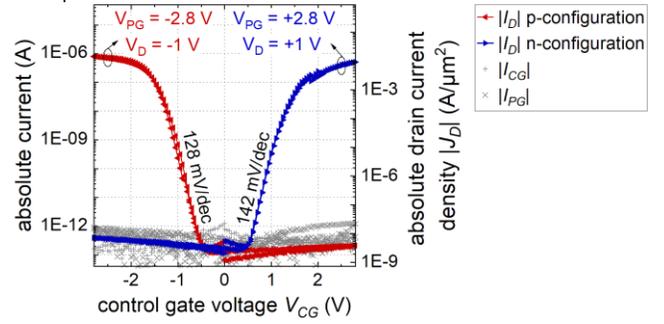

Fig. 2. Schematic of the RFET band structure for different voltage settings. The polarity of the voltages at drain and program gate (PG) determines the p- or n-configuration. By tuning the control gate (CG) voltage to the same polarity, carriers are injected at the source by tunneling through the Schottky junction.

Fig. 4. Highly symmetric transfer characteristics of the RFET for a drain voltage of 1 V (n-configuration) and -1 V (p-configuration), both in a double sweep. Equal on- and off-currents as well as a high on/off ratio in both configurations are achieved. Gate charging and leakage currents (grey) remain low even for high reverse $V_{CG}$.

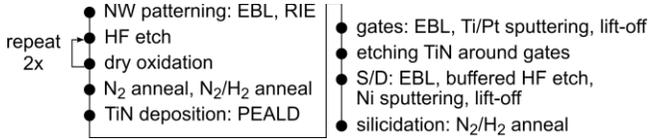

Fig. 3. Process flow for transistor fabrication from nanowire (NW) creation to source/drain (S/D) contact formation.

The current densities of the device in this work are the highest reported so far for top-down fabricated RFETs. Furthermore, the on/off ratios are very high and the transfer characteristics are almost symmetric for both configurations.

## II. DEVICE FABRICATION

The RFET in this work is fabricated from a commercial SOI wafer with a 20 nm thick, (001) oriented and lightly p-doped ($10^{15}$ cm$^{-3}$) device layer on top of a 100 nm thick buried $SiO_2$ layer. Fig. 3 gives an overview of the process flow. A 3.98 µm long nanowire channel is created in <110> direction by using an electron beam lithography (EBL) defined hydrogen silsesquioxane (HSQ) pattern as hard mask for a reactive ion etching process [31]. The etching employs $SF_6$, $CHF_3$ and $O_2$ at a ratio of 15:6:5 and a pressure of 0.1 Torr.

After removing the residual HSQ and native oxide by a dip in HF, the $SiO_2$ gate dielectric is formed by rapid thermal annealing at 875°C. The dry oxidation of narrow Si nanowires is known to be self-limiting below approximately 950°C, making it highly reliable [32]. By interrupting the oxidation multiple times to apply HF etches, a well-controlled diameter reduction and a partial underetch of the nanowire are achieved. The cross section TEM in Fig. 1b reveals a channel height of 3.5 nm, a width of approx. 12–16 nm (16 nm assumed for Fig. 4 and 5) and an oxide shell thickness of circa 6.5 nm. After the last oxidation, the chip is annealed at $N_2$ and $N_2/H_2$ (9:1) atmosphere at 875°C and 450°C, respectively, to reduce defects in the $SiO_2$ and at its interface to the silicon.

Subsequently, the wafer is coated with 12 nm of TiN by plasma enhanced atomic layer deposition (PEALD) based on a $TiCl_4$ precursor and $N_2/H_2$ as co-reactant at a temperature of 250°C. X-ray photoelectron spectroscopy reveals a Ti:N ratio of 47:53 and capacitance-voltage measurements a work function of 4.81 eV of the TiN. Further, a stack of 3 nm Ti and 37 nm Pt is deposited by sputtering and structured by means of an EBL-based lift-off. The resulting Ω-shaped gate architecture can be seen in Fig. 1b. TiN is then etched around the gates by a mixture of $H_2O$, ammonia water and $H_2O_2$.

Source/drain contact areas are defined by another EBL. The oxide shell is locally removed by a dip in $NH_4F$ buffered HF and 40 nm Ni are sputtered. After removing undesired Ni by a lift-off, a rapid thermal anneal at 450°C in forming gas atmosphere is performed to intrude Ni into the wire to create atomically sharp $NiSi_2$/Si Schottky junctions below the gates [31].

## III. DEVICE CHARACTERISTICS

Both Schottky junctions of the RFET are gated individually by a so-called program gate (PG) at the drain and a control gate (CG) at the source side (see Fig. 2). $NiSi_2$ has a work function close to middle of the band gap of Si, with a slightly larger SB for electrons [33]. For electrical characterization, the substrate and the source terminal are grounded. By changing the polarity of the drain voltage $V_D$ and program gate voltage $V_{PG}$, the RFET can be toggled between p- and n-configuration by blocking undesired carrier injection from the drain. For $V_{PG} \geq V_D > 0$ V, hole injection from the drain is blocked by the large energy barrier, so the RFET is in n-configuration. To turn the transistor on, a sufficiently high voltage is applied at the CG. This induces Fowler-Nordheim tunneling of electrons through the source-sided Schottky junction. For p-configuration the polarities of all voltages are simply inverted so that holes are injected from the source side in the on-state.

For complementary logic applications, a symmetry of both configurations in terms of $V_T$ and on-current is desired to ensure that circuit timings are invariant of the polarity of the active RFETs. This is of special relevance for protecting circuits against side channel attacks [6]. Therefore, the intrinsically higher SB for electrons than for holes must be compensated. The currents can be aligned by shifting the transfer curves towards negative gate voltages by means of the gate metal work function or fixed oxide charges. Yet, this would result in an undesirably increased off-current (at $V_{CG} = 0$ V) for the n-configuration. Simulations revealed that also the compressive stress generated by the thermally grown oxide shell of a nanowire can induce changes in the band structure. This results in a reduced barrier height and effective tunneling mass of electrons which increases the on-current in n-configuration [34]. For holes the compressive stress has opposite effects so that p- and n-current can be equilibrated for a suitable oxidation time. Stress by the metal gate may also contribute to this current equalization [34]. In any case, the omega-geometry applies the stress of the gate stack almost ideally from all sides to the channel.

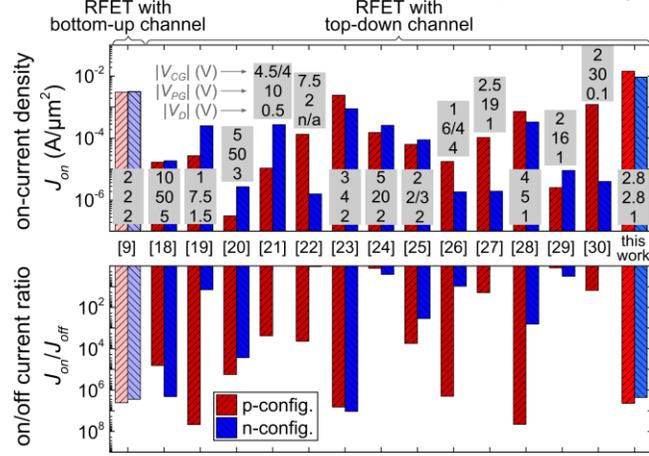

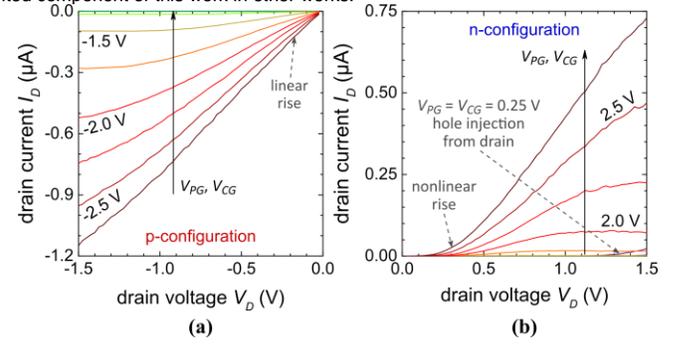

Fig. 6. (a) Output characteristics of the RFET for p-configuration. Gate voltages are changed in 250 mV steps. (b) Output characteristics for n-configuration. Gate voltages are changed in 250 mV steps.

Fig. 5. Performance comparison of one bottom-up and 14 top-down fabricated Si-based RFETs regarding maximal on-current density and corresponding on/off current ratio for p- (red) and n-configuration (blue). Operation points are at room temperature and selected for equal absolute drain, CG and PG voltages if available in p-/n-configuration (boxes in upper plot). Off-states refer to $V_{CG} = 0$ V. [9] has an approx. 7.5 nm wide nanowire channel according to the author. For [30] on-state currents in n-configuration are estimated. For [21] and [30] no on/off current ratios are extractable for n-configuration.

Transfer characteristics in Fig. 4 show that the on-current in p-configuration (803 nA) is in fact only factor 1.6 larger than in n-configuration (516 nA). Note that this symmetry is achieved for equal absolute gate voltages of $|V_{CG}| = |V_{PG}| = 2.8$ V. Considering the narrow cross-section (circa 56 nm$^2$), a remarkable current density of up to 14.3 mA/µm$^2$ is achieved. Higher total currents can be achieved by stacking multiple nanowires and reducing the channel length [2], [23], [35]. For a stack of multiple 10 nm wide nanowires and gate spacings and widths of both 10 nm, on-currents per width are simulated to reach that of low-standby-power CMOS transistors of the 5 nm node [35], [36].

Our RFET further offers a low hysteresis and a high on/off current ratio for both configurations when considering $V_{CG} = 0$ V as off-state. The off-current even remains low for high reverse CG voltages because the gate leakage currents are very low and – in contrast to SB-MOSFETs – a reverse carrier injection from drain is effectively blocked by the PG.

Fig. 5 compares the results to prior art Si-based RFETs based on uniform benchmark criteria to accommodate for their versatile architectures and measurement settings. The gate at the drain is always considered as the PG while the independent gate at the source or in the middle of the channel is the CG. The operation points are chosen for equal absolute voltages in both configurations whenever possible, i.e. for $V_{Dn} = -V_{Dp}$, $V_{PGn} = -V_{PGp}$ and $V_{CGn} = -V_{CGp}$. Note that if p-configuration has been unusually measured at positive $V_D$, all voltages refer to this as the real ground potential. Then $V_S = -V_D$ at source is the real drain potential. Amongst the applicable operation points, the maximal current densities (on-state) and the current at $V_{CG} = 0$ V (off-state) are extracted. This work presents the first top-down fabricated Si-based RFET to exceed the high on-current densities of the bottom-up RFET [9]. Coincidently, the on-currents are similar and the on/off ratio is very high for both configurations.

The subthreshold swing $SS$ reaches 128 mV/dec for p- and 142 mV/dec for n-configuration. Lower values down to 63 mV/dec (6 mV/dec with impact ionization) for top-down fabricated RFETs have only been reported when the CG controls the middle of the channel to create potential barriers for already injected carriers [23], [26], [28]. In this work by contrast the CG is at the Schottky junction and directly tunes the injection of carriers. This gating approach has always resulted in $SS \geq 150$ mV/dec in silicon-based RFETs [8]–[10], [25].

Output characteristics are presented in Fig. 6 for simultaneously varied potentials at CG and PG. The current saturates when the carriers can leave the channel at drain without a barrier. A deferred and non-linear rise of negative curvature is notable in the n-configuration being typical for Schottky-junction-based devices [37], [33]. It probably originates from the tunneling barrier for electrons at the drain caused by the $V_{PG}$ to $V_D$ potential difference. As $V_D$ rises, the barrier width decreases, leading to the initially exponential increase in tunneling transmission. By contrast, due to the lower NiSi$_2$-Si Schottky barrier for holes than for electrons, the holes can already tunnel with high probability in p-configuration even for low $V_D$. With increasing $V_D$, the thermal emission increases so that $I_D$ rises linearly. For low gate voltages in n-configuration a slightly risen current at high $V_D$ can be observed. It originates from the reverse injection of holes from the drain due to the rising gate-drain potential difference.

## IV. SUMMARY

A reconfigurable FET with symmetric transfer characteristics is fabricated in a top-down manner. It exhibits minimal hysteresis, high on- and low off-currents for both, n- and p-configuration. This shows the importance of a narrow nanowire channel and encasing gate especially for RFETs.